\documentclass [12pt]{article}
\usepackage {graphicx}
\usepackage {longtable}

\begin{document}
\begin{center}
\textbf{CONFORMAL GEOMETRODYNAMICS: EXACT NONSTATIONARY SPHERICALLY 
SYMMETRIC SOLUTIONS}
\end{center}

\bigskip

\begin{center}
\textbf{M.V. Gorbatenko}
\end{center}

\bigskip

\begin{center}
Russian Federal Nuclear Center - All-Russian Research Institute of 
Experimental Physics, Sarov, Nizhnii Novgorod region, 
\end{center}

\begin{center}
E-mail: {gorbatenko@vniief.ru}
\end{center}

\bigskip

\begin{center}
\textbf{Abstract}
\end{center}

\bigskip

A nonstationary spherically symmetric problem for conformal geometrodynamics 
equations is considered and general exact solutions in quadratures are 
obtained. Involvement of Weyl degrees of freedom allows us to consider the 
problem with arbitrary initial data, as for the conformal geometrodynamics 
equations the Cauchy problem is set up without connections to initial data. 
The results of this paper are not confined with the framework of the 
perturbation theory and open up new avenues for study of the process of 
space-time singularity evolution in time.

\bigskip

\section*{1. Introduction. CGD equations}

\bigskip

It is well known that the general relativity equations for the empty 
Riemannian space can be used to describe dynamic evolution only of those 
spaces, which on the initial space-like hypersurface satisfy four 
connections. Given a spherically symmetric (SphS) problem, because of the 
connections the general relativity equations cannot be used to consider the 
space evolution with arbitrary spherically symmetric distribution of the 
metric tensor components at the initial time. 

A way out of this situation is sought for by different investigators in 
different directions: some of them through inclusion of Riemann-tensor 
quadratic terms in the general relativity equations (see, e.g., [1]) while 
the others through inclusion of field fluctuations breaking the spherical 
symmetry and carrying away particle multipole moments near the event horizon 
in the form of Hawking heat emission (see, e.g., [2]). The typically used 
method for studying field configurations (including the singular ones) 
therewith involves the notion of the observer, which is a trial material 
body fit out with frame attributes and moving freely in a space under 
consideration. That is the method is used in which the field configuration 
is considered as the one that has already appeared, its appearance history 
is not taken into account. One more way-out direction is pursued through the 
search for the exact general relativity equation solution for the body 
generating the Schwarzschild field and the particle perturbing the field 
irrespective of how small the difference from the sphericity would be (see, 
e.g., [3]).

It may turn out that all or some field configurations considered in all 
these approaches can never be realized in principle. This can happen, for 
example, when the dynamic equations lead to appearance and development of 
discontinuity surfaces near potential points of appearance of the 
singularities. In these cases the discontinuity surfaces in essence can play 
the role of field configuration regularizers impeding the singularity 
formation. 

However, the mentioned mechanism of counteraction to the appearance of 
singularities a fortiori cannot appear when the general relativity equations 
for empty space are used for the dynamics equations. These equations are 
known to admit only weak discontinuities (discontinuities only of second 
normal derivatives of the metric component) and only on light-like surfaces.

As a candidate for more general dynamic equation of particular interest are 
the conformal geometrodynamics (CGD) equations derived in [4] (without 
lambda term) and in [5] (with lambda term). The CGD equations describe Weyl 
empty space dynamics and, as shown in [6], admit setting up the Cauchy 
problem without connections to initial data. It is this property of the CGD 
equations that allows us to consider the time evolution of Riemannian space 
with arbitrary initial data. 

In this paper, for simplicity we restrict our consideration to the CGD 
equations without lambda term. According to [4], in this case the CGD 
equations are given by

\begin{equation}
\label{eq1}
R_{\alpha \beta}  - \frac{{1}}{{2}}g_{\alpha \beta}  R = T_{\alpha \beta}  
,
\end{equation}

\noindent
where

\begin{equation}
\label{eq2}
T_{\alpha \beta}  = - 2A_{\alpha}  A_{\beta}  - g_{\alpha \beta}  A^{2} - 
2g_{\alpha \beta}  A^{\nu} _{;\nu}  + A_{\alpha ;\beta}  + A_{\beta ;\alpha 
} .
\end{equation}

As it follows from (\ref{eq2}), alongside metric tensor $g_{\alpha \beta}  $ vector 
$A_{\alpha}  $appears in $T_{\alpha \beta}  $. CGD equations (\ref{eq1}) with tensor 
$T_{\alpha \beta}  $of form (\ref{eq2}) differ from the general relativity equations 
for empty space

\begin{equation}
\label{eq3}
R_{\alpha \beta}  - \frac{{1}}{{2}}g_{\alpha \beta}  R = 0
\end{equation}

\noindent
in two points: first, in the presence of the nonzero energy-momentum tensor 
$T_{\alpha \beta}  $ in the right-hand side; second, in a specific structure 
of the tensor. As a consequence of the structure the CGD equations possess 
invariance under conformal transformations

\begin{equation}
\label{eq4}
\left. {\begin{array}{l}
 {g_{\alpha \beta}  \to {g}'_{\alpha \beta}  = g_{\alpha \beta}  \cdot 
e^{2\sigma} } \\ 
 {A_{\alpha}  \to {A}'_{\alpha}  = A_{\alpha}  - \sigma _{,\alpha} }  \\ 
 \end{array}}  \right\}.
\end{equation}

Here $\sigma $ is an arbitrary scalar function.

The first difference ($T_{\alpha \beta}  \ne 0$) leads to the fact that the 
dynamic equations begin to admit existence of space-like discontinuity 
surfaces, for example, in the form of shock waves. At the same time, the 
dynamic equations used by us describe Weyl empty space dynamics. Thus, the 
CGD equations are in essence the implementation of Wheeler's idea about the 
purely geometrodynamic description of matter, but in the Weyl rather than 
Riemannian space. The second difference (the invariance under conformal 
transformations) leads to uniqueness of the structure of the tensor 
$T_{\alpha \beta}  $.

We now turn to the discussion of the SphS problem.

First of all recall that in the general relativity the statement known as 
Israel's theorem (see [7]) has been proved, according to which (see, e.g., 
[8]-[10]) the solution to the SphS problem in flat asymptotics at infinity 
is automatically static. Besides, in this problem nothing except 
Schwarzschild solution or its equivalent (with an accuracy to coordinate 
transformation on some map) can be obtained whatsoever. 

In the case of the CGD equations the SphS problem possesses a great variety 
of solutions. The static SphS solutions to the CGD equations complemented 
with a so-called lambda term are discussed in detail in ref. [11]. The 
solutions are shown to include three branches, each of which is determined 
by five integration constants. In this paper we will take up the research 
into another wide class of the SphS solutions to the CGD equations, i.e. the 
nonstationary SphS solutions. For the general relativity analogs of the 
nonstationary SphS solutions can be obtained only with nonzero 
energy-momentum tensors over the entire space. From the standpoint of the 
general relativity CGD equations (\ref{eq1}) are an alternative of the general 
relativity equations with the energy-momentum tensor of some matter.

\section*{2. The spherically symmetric problem for the CGD equations}

\bigskip

\subsection*{2.1. The general form of the metric and Weyl vector in the SphS problem}

\bigskip

Here we will not derive the spherically symmetric metric. The derivations 
can be found in many general relativity monographs, see, e.g., [8]. The 
metric that will be hereinafter referred to as spherically symmetric is 
given by

\begin{equation}
\label{eq5}
ds^{2} = - e^{\gamma}  \cdot dt^{2} + e^{\alpha}  \cdot dx^{2} + e^{\beta}  
\cdot \left[ {d\theta ^{2} + sin^{2}\theta \cdot d\varphi ^{2}} \right].
\end{equation}

Besides assumptions of the metric, certain assumptions of the structure of 
vector $A_{\alpha}  $should be made when solving the CGD equations. We 
assume that of 4 independent components of vector $A_{\alpha}  $ in case of 
SphS as few as two remain, i.e.

\begin{equation}
\label{eq6}
A_{\alpha}  = \left( {A_{0} ,A_{1} ,A_{2} ,A_{3}}  \right)\quad \to \quad 
A_{\alpha}  = \left( {\varphi ,f,0,0} \right).
\end{equation}

All the five introduced functions $\alpha ,\;\beta ,\;\gamma $, $\varphi 
,\;f$ are functions of time $t$ and radial variable $x$. The metric in form 
(\ref{eq5}) is considered in detail in ref. [8], which also presents expressions for 
Christoffel symbols that will be used in what follows as well as for Riemann 
tensor components.

When reasoning from the metric in form (\ref{eq5}), it can be noticed that the 
radial variable is determined not uniquely by this form. Performing the 
transformation of the radial variable, we notice that the metric form 
remains unchanged, but relations between components $g_{11} $ and $g_{22} $ 
do change. The radial variable can be selected so, that functions $\alpha $ 
and $\beta $ coincide,

\begin{equation}
\label{eq7}
\alpha = \beta .
\end{equation}

In what follows as assume that the choice of the radial variable is made 
just in this way, i.e. so that relation (\ref{eq7}) hold.

The coordinate transformation spoken about above can be performed both in 
the general relativity and in CGD. But in the case of the CGD equations the 
``freedom'' of transformation of functions $\alpha ,\;\beta ,\;\gamma $, 
$\varphi ,\;f$ is not limited to the possibility to attain the fulfillment 
of relation (\ref{eq7}). One more transformation can be performed, viz. the 
conformal transformation with function $\sigma $ [see (\ref{eq4})]

\begin{equation}
\label{eq8}
\sigma = - \frac{{1}}{{2}}\beta .
\end{equation}

Upon the two above transformations for the CGD equations the metric form 
will include only one function $\gamma $,

\begin{equation}
\label{eq9}
ds^{2} = - e^{\gamma}  \cdot dt^{2} + dx^{2} + \left[ {d\theta ^{2} + 
sin^{2}\theta \cdot d\varphi ^{2}} \right],
\end{equation}

\noindent
with the form of vector $A_{\alpha}  $ (\ref{eq6}) remaining unchanged. 

Thus, in what follows we take $\alpha = 0,\;\beta = 0$ in CGD equations, so 
that of five functions $\alpha ,\;\beta ,\;\gamma $, $\varphi ,\;f$ only 
three remain: $\gamma $, $\varphi ,\;f$. After that the CGD equations for 
the three functions are written in the following form:

\begin{equation}
\label{eq10}
 - 1 = e^{ - \gamma}  \cdot 3\varphi ^{2} - f^{2} - 2{f}',
\end{equation}

\begin{equation}
\label{eq11}
 - 1 = e^{ - \gamma}  \cdot \left[ {\varphi ^{2} + 2\dot {\varphi}  - \dot 
{\gamma} \varphi}  \right] - 3f^{2} - {\gamma} 'f,
\end{equation}

\begin{equation}
\label{eq12}
0 = - 2\varphi f + {\varphi} ' + \dot {f} - {\gamma} '\varphi ,
\end{equation}

\begin{equation}
\label{eq13}
\frac{{{\gamma} ''}}{{2}} + \frac{{{\gamma} '^{2}}}{{4}} = e^{ - \gamma}  
\cdot \left[ {\varphi ^{2} + 2\dot {\varphi}  - \dot {\gamma} \varphi}  
\right] - f^{2} - 2{f}' - {\gamma} 'f.
\end{equation}

Instead of function $\gamma $ it is convenient to introduce a new function 
$C$ defined as

\begin{equation}
\label{eq14}
C = \sqrt { - g} = \sqrt {e^{\gamma} } = e^{\gamma /2}.
\end{equation}

We obtain equations (\ref{eq10})-(\ref{eq13}) in the new form:

\begin{equation}
\label{eq15}
 - 1 = 3\frac{{\varphi ^{2}}}{{C^{2}}} - f^{2} - 2{f}'
\end{equation}

\begin{equation}
\label{eq16}
 - 1 = \frac{{1}}{{C^{2}}} \cdot \left[ {\varphi ^{2} + 2\dot {\varphi}  - 
2\varphi \frac{{\dot {C}}}{{C}}} \right] - 3f^{2} - 2f\frac{{{C}'}}{{C}}
\end{equation}

\begin{equation}
\label{eq17}
0 = - 2\varphi f + {\varphi} ' + \dot {f} - 2\varphi \frac{{{C}'}}{{C}}
\end{equation}

\begin{equation}
\label{eq18}
\frac{{{C}''}}{{C}} = \frac{{1}}{{C^{2}}} \cdot \left[ {\varphi ^{2} + 2\dot 
{\varphi}  - 2\varphi \frac{{\dot {C}}}{{C}}} \right] - f^{2} - 2{f}' - 
2f\frac{{{C}'}}{{C}}
\end{equation}

Having extracted (\ref{eq15}) and (\ref{eq16}) from (\ref{eq18}), we arrive at:

\begin{equation}
\label{eq19}
\frac{{{C}''}}{{C}} = - 2 - 3\frac{{\varphi ^{2}}}{{C^{2}}} + 3f^{2}.
\end{equation}

Thus, equation (\ref{eq19}) can be considered instead of (\ref{eq18}) in what follows. 

Ref. [12] shows that from the CGD equations of form (\ref{eq1}) with energy-momentum 
tensor (\ref{eq2}) equation

\begin{equation}
\label{eq20}
F^{\alpha \sigma} _{;\sigma}  = \frac{{1}}{{\sqrt { - g}} }\frac{{\partial 
}}{{\partial x^{\sigma} }}\left( {\sqrt { - g} \;F^{\alpha \sigma} } \right) 
= 0.
\end{equation}

From (\ref{eq20}) it follows that

\begin{equation}
\label{eq21}
\dot {f} = {\varphi} ' - C \cdot \phi _{0} .
\end{equation}

Emphasize that relation (\ref{eq21}) follows from CGD equations (\ref{eq15})-(\ref{eq17}), (\ref{eq19}). The 
particular attention paid to relation (\ref{eq21}) is due to its simplicity.

Rearrange equations (\ref{eq15})-(\ref{eq17}), (\ref{eq19}), having made one more function 
substitution,

\begin{equation}
\label{eq22}
z \equiv \frac{{\varphi} }{{C}},
\end{equation}

\noindent
and using relation (\ref{eq21}) to eliminate $\dot {f}$. Equation (\ref{eq15}) will become

\begin{equation}
\label{eq23}
{f}' = \frac{{1}}{{2}} + \frac{{3}}{{2}}z^{2} - \frac{{1}}{{2}}f^{2}.
\end{equation}

Equation (\ref{eq16}) will assume form

\begin{equation}
\label{eq24}
\frac{{\dot {z}}}{{C}} = - \frac{{1}}{{2}} - \frac{{1}}{{2}}z^{2} + 
\frac{{3}}{{2}}f^{2} + f \cdot \left( {\frac{{{C}'}}{{C}}} \right).
\end{equation}

Equation (\ref{eq17}) will take form

\begin{equation}
\label{eq25}
{z}' = \frac{{\phi _{0}} }{{2}} + fz.
\end{equation}

Finally, equation (\ref{eq19}) will become

\begin{equation}
\label{eq26}
\frac{{{C}''}}{{C}} = - 2 + 3T,
\end{equation}

\noindent
where

\begin{equation}
\label{eq27}
T \equiv f^{2} - z^{2}.
\end{equation}

Using definition (\ref{eq27}) for function $T$ and formula (\ref{eq21}) for $\dot {f}$, 
equation (\ref{eq24}) can be written as

\begin{equation}
\label{eq28}
\frac{{\dot {T}}}{{C}} = \left( {z - \phi _{0} f - zT} \right).
\end{equation}

Equations (\ref{eq21}), (\ref{eq23})-(\ref{eq26}) [or their equivalent equations (\ref{eq21}), (\ref{eq23}), 
(\ref{eq25})-(\ref{eq28})] just make up the equation system, to the solution of which we 
will pass on right now.

\subsection*{2.2. Determination of desired functions}

\bigskip

By direct verification we make sure that from equations (\ref{eq23}), (\ref{eq25}) relation

\begin{equation}
\label{eq29}
\left( {z - \phi _{0} f - zT} \right)^{\prime}  = 0.
\end{equation}

\noindent
follows. Thus, there is relation among functions $z,f,T$:

\begin{equation}
\label{eq30}
z - \phi _{0} f - zT = F_{1} \left( {t} \right),
\end{equation}

\noindent
where $F_{1} \left( {t} \right)$ is an arbitrary function of time. Relation 
(\ref{eq30}) is in essence the first integral of equations (\ref{eq23}), (\ref{eq25}). And it can be 
used to determine the entire set of the desired functions. It is convenient 
to perform the further manipulations using the above introduced functions 
$z$ and $T$.

Substitute (\ref{eq30}) into (\ref{eq28}).

\begin{equation}
\label{eq31}
\dot {T} = F_{1} \left( {t} \right) \cdot C.
\end{equation}

Hence it follows that to obtain the stationary SphS solution function $F_{1} 
\left( {t} \right)$ should be set identically zero. In other words:

\begin{equation}
\label{eq32}
Stationary\quad solution\quad \to \quad F_{1} \left( {t} \right) \equiv 0.
\end{equation}

We are seeking the nonstationary solution, therefore hereinafter we assume 
that $F_{1} \left( {t} \right) \ne 0$. Under this assumption from equations 
(\ref{eq31}) and (\ref{eq26}) it follows that

\[
\frac{{{\dot {T}}''}}{{\dot {T}}} = - 2 + 3T,
\]

\noindent
i.e.

\[
{\dot {T}}'' + \left( {2 - 3T} \right)\dot {T} = 0.
\]

If this equation is written as

\[
\frac{{\partial} }{{\partial t}}\left[ {{T}'' + 2T - \frac{{3}}{{2}}T^{2}} 
\right] = 0,
\]

\noindent
then it will follow from this that

\begin{equation}
\label{eq33}
{T}'' + 2T - \frac{{3}}{{2}}T^{2} = F_{2} \left( {x} \right),
\end{equation}

\noindent
where $F_{2} \left( {x} \right)$ is some function of radial variable. 
Determine the function. From the definition of $T$ it follows that

\[
{T}' = 2\left( {f{f}' - z{z}'} \right),
\]

\[
{T}'' = 2\left( {f{f}'' + {f}'^{2} - z{z}'' - {z}'^{2}} \right).
\]

The expressions for ${f}'$ and ${z}'$ are determined by equations (\ref{eq23}) and 
(\ref{eq25}), respectively. Second derivatives ${f}''$, ${z}''$ will be determined 
from those same equations by straightforward differentiation. Upon the 
substitution of the expressions for derivatives in (\ref{eq33}) we obtain:

\begin{equation}
\label{eq34}
F_{2} \left( {x} \right) = \frac{{1}}{{2}}\left( {1 - \phi _{0}^{2}}  
\right).
\end{equation}

Here $\phi _{0} $ is the same constant which appeared in relation (\ref{eq21}). 
Substitute (\ref{eq34}) into (\ref{eq33}).

\begin{equation}
\label{eq35}
{T}'' = - 2T + \frac{{3}}{{2}}T^{2} + \frac{{1}}{{2}}\left( {1 - \phi 
_{0}^{2}}  \right).
\end{equation}

Multiply resultant equation (\ref{eq35}) by $2{T}'$.

\begin{equation}
\label{eq36}
2{T}''{T}' = \left[ { - 4T + 3T^{2} + \left( {1 - \phi _{0}^{2}}  \right)} 
\right]{T}'.
\end{equation}

Equation (\ref{eq36}) can be integrated.

\begin{equation}
\label{eq37}
\left( {{T}'} \right)^{2} = - 2T^{2} + T^{3} + \left( {1 - \phi _{0}^{2}}  
\right)T + F_{3} \left( {t} \right).
\end{equation}

Here $F_{3} \left( {t} \right)$ is some function of time. The function can 
be determined. To do this replace ${T}'$ in (\ref{eq37}) with expressions in terms 
of variables ${f}'$ and ${z}'$ and for the derivatives use equations (\ref{eq23}), 
(\ref{eq25}). As a result we determine that

\begin{equation}
\label{eq38}
F_{3} \left( {t} \right) = F_{1}^{2} \left( {t} \right).
\end{equation}

Substitute the determined expression for $F_{3} \left( {t} \right)$ into 
equation (\ref{eq37}) and arrive at equation

\begin{equation}
\label{eq39}
\frac{{dT}}{{\sqrt { - 2T^{2} + T^{3} + \left( {1 - \phi _{0}^{2}}  \right)T 
+ F_{3} \left( {t} \right)}} } = dx.
\end{equation}

Integrate.

\begin{equation}
\label{eq40}
x + F_{4} \left( {t} \right) = \int {\frac{{dT}}{{\sqrt {T^{3} - 2T^{2} + 
\left( {1 - \phi _{0}^{2}}  \right)T + F_{3} \left( {t} \right)}} }} .
\end{equation}

The integral appearing on the right is expressed in terms of the well-known 
doubly periodic Weierstrass function (see, e.g., [13]). In the function one 
period is real, while the other is purely imaginary. The periods depend on 
$F_{1} \left( {t} \right)$ and $\phi _{0} $. 

If we transfer to integration variable

\begin{equation}
\label{eq41}
s = T - \frac{{2}}{{3}},
\end{equation}

\noindent
then relation (\ref{eq40}) is reduced to form

\begin{equation}
\label{eq42}
x + F_{4} \left( {t} \right) = - 2\int\limits_{T - \frac{{2}}{{3}}}^{\infty 
} {\frac{{ds}}{{\sqrt {4s^{3} - g_{2} s - g_{3}} } }} .
\end{equation}

Here:

\begin{equation}
\label{eq43}
g_{2} = \frac{{4}}{{3}} + 4\phi _{0}^{2} ;\quad g_{3} = - \frac{{8}}{{27}} - 
4F_{1}^{2} .
\end{equation}

In the standard notation $\wp $ for the Weierstrass function we have:

\begin{equation}
\label{eq44}
T - \frac{{2}}{{3}} = \wp \left( { - \frac{{x + F_{4} \left( {t} 
\right)}}{{2}};g_{2} ;g_{3}}  \right).
\end{equation}

Function $F_{4} \left( {t} \right)$ appearing in $\wp $, as well as function 
$F_{1} \left( {t} \right)$ and constant $\phi _{0} $ appearing in the 
expressions for $g_{2} $ and $g_{3} $ are arbitrary. 

Formula (\ref{eq44}) gives the answer to the question of the general solution for 
function $T$. The expression for function $C$ is determined using formula 
(\ref{eq31}). It remains to determine functions $z$ and $f$. For this purpose we 
consider relation (\ref{eq27}) and first integral relations (\ref{eq30}) as two algebraic 
equations for determination of functions $z$ and $f$. Solving these 
relations, we determine that:

\begin{equation}
\label{eq45}
z = \frac{{F_{1} \cdot \left( {1 - T} \right) + \eta \phi _{0} \cdot \sqrt 
{T^{3} - 2T^{2} + \left( {1 - \phi _{0}^{2}}  \right)T + F_{1}^{2}}  
}}{{\left[ {\left( {1 - T} \right)^{2} - \phi _{0}^{2}}  \right]}};
\end{equation}

\begin{equation}
\label{eq46}
f = \frac{{\phi _{0} F_{1} + \eta \left( {1 - T} \right) \cdot \sqrt {T^{3} 
- 2T^{2} + \left( {1 - \phi _{0}^{2}}  \right)T + F_{1}^{2}} } }{{\left[ 
{\left( {1 - T} \right)^{2} - \phi _{0}^{2}}  \right]}}.
\end{equation}

In these expressions functions $z$ and $f$ are expressed only in terms of 
$T$, arbitrary time function $F_{1} \left( {t} \right)$ and some constant 
$\phi _{0} $. In these expressions $\eta $ takes on the values of either 
$\eta = 1$ or $\eta = - 1$. 

Substituting the resultant expressions for functions $z$ and $f$ into the 
initial dynamic equations, we can see that these equations get transformed 
into identities, if function $T$ is determined by relation (\ref{eq44}) and function 
$C$ is calculated using formula (\ref{eq31}). These transformations are quite 
straightforward, but cumbersome, therefore we will not present them here.

Thus, we have found the general solutions for all the functions determining 
the nonstationary SphS solution.

\section*{3. Discussion}

\bigskip

This paper has derived the general expressions for functions $T\left( {t,x} 
\right)$, $C\left( {t,x} \right)$, $z\left( {t,x} \right)$, $f\left( {t,x} 
\right)$ appearing in the consideration of the nonstationary SphS problem. 
The expressions for the functions are determined by formulas (\ref{eq44}), (\ref{eq31}), 
(\ref{eq45}), (\ref{eq46}), respectively. 

Two special solutions have been constructed\footnote{ To be published in 
VANT-TPF.} . One of them can be associated with a train of spherically 
symmetric waves converging to center or diverging from it. In either case 
conformal invariant $\left( {F_{\alpha \beta}  F_{\mu \nu}  g^{\alpha \mu 
}g^{\beta \nu} } \right)$ is nonsingular and nonzero, so that the solutions 
a fortiori do not refer to the category of conformally flat ones. In either 
solution this invariant has a finite value.

In connection with the results note that the solutions with spherical 
symmetry for an empty space refer to simplest exact solutions both in the 
general relativity and CGD. In the case of the general relativity the SphS 
solution is unique and reduces to Schwarzschild solution. In the case of CGD 
the SphS solution set consists of two classes: stationary and nonstationary 
solutions. The former are analogs of the Schwarzschild solution. As for the 
nonstationary solutions studied in this paper, in the general relativity 
there are no analogs of them for the case of empty Riemannian spaces. At the 
same time it is clear that it is the nonstationary solutions that may be of 
the greatest interest for the study of the phenomena, such as supernova 
explosions, Hawking vaporization near event horizons, etc. This supposition 
can prove valid, if the above phenomena can appear only evolutionarily, i.e. 
as a result of completion of some phase of the nonstationary solution 
evolution. Therefore the resort to the nonstationary SphS solutions of the 
CGD equations opens up new avenues for studying the above-mentioned 
phenomena. 

The appearance of the new class of the nonstationary SphS solutions in the 
transition from the general relativity to the CGD equations is not 
occasional. The reason is that in the case of the CGD equations the Cauchy 
problem is set up without connections to Cauchy data on the initial 
space-like hypersurface. The disappearance of four connections to Cauchy 
data that take place in the case of the general relativity equations 
unfreezes the degrees of freedom that just make the nonstationary SphS 
solution appearance possible. In other words, the CGD equations can be used 
to numerically calculate the metric and Weyl vector evolution under 
arbitrary initial and boundary conditions. This CGD equation property seems 
to us basically important and serves, in our opinion, a weighty argument in 
favor of taking into consideration the Weyl degrees of freedom of space-time 
in cosmological studies. 

In conclusion note that using the results of this paper one may try to 
verify the unbounded energy cumulation instability hypothesis advanced in 
ref. [14]. In our dynamic equations the cumulation bounding mechanism is 
associated with the Weyl degrees and appearance of a discontinuity surface 
in geometrodynamic continuum with the equations of state generated by the 
equations themselves.

\bigskip

\section*{References}

\bigskip

[1] R.Bach and H.Weyl. Z. \textbf{13}, 134 (1920).

[2] S.W.Hawking. Commun. Math. Phys. \textbf{43}, 199 (1975).

[3] L.Herrera. \textit{Possible way out of the Hawking paradox: Erasing the 
information at the horizon.} arXiv: 0709.4674v1 [gr-qc] 28 Sep 2007.

[4] M.V.Gorbatenko, A.V.Pushkin. \textit{Dynamics of the linear affine 
connectedness space and conformally invariant extension of Einstein 
equations.} // VANT-TPF$^{} $\footnote{ Hereinafter by VANT-TPF$^{} $is 
meant Colelction ``Voprosy Atomnoy Nauki i Tekhniki. Series: Teoreticheskaya 
i Prikladnaya Fizika ''.} . \textbf{2/2}, 40 (1984)

[5] Yu.A.Romanov. \textit{Affine connectedness space dynamics}. // VANT-TPF 
. \textbf{3}, 55 (1996).

[6] M.V.Gorbatenko, Yu.A.Romanov. \textit{Cauchy problem for equations 
describing the affine connectedness space dynamics}. // VANT-TPF . 
\textbf{2}, 34 (1997).

[7] W.Israel. Phys. Rev. \textbf{164}, 1776 (1967).

[8] J.Sing. \textit{The general relativity}. Moscow. Inostrannaya Literatura 
Publishers (1963).

[9] L.D.Landau, E.M.Lifshits. \textit{The field theory}. Moscow. Nauka 
Publishers (1988).

[10] A.Z.Petrov. \textit{New methods in the general relativity}. Moscow. 
Nauka Publishers (1966).

[11] M.V.Gorbatenko. \textit{Discontinuous solutions in conformally 
invariant geometrodynamics}. // Advances in Mathematics Research. Vol. 6. 
Editors:\textbf{} Oyibo, Gabriel. Nova Science Publishers, Inc. (2005).

[12] M.V.Gorbatenko, A.V.Pushkin. \textit{Conformally Invariant 
Generalization of Einstein Equation and the Causality Principle}. // General 
Relativity and Gravitation. \textbf{34}, No. 2, 175-188 (2002). 

[13] E.T.Wittecker, J.N.Watson. \textit{Course of modern analysis.} V. II. 
Moscow. GIFML Publishers (1963).

[14] E.I.Zababakhin. Pis'ma v ZhETF. \textbf{30}, 97 (1979).

\end{document}